# Deriving a comprehensive dataset of optical constants for metal halide perovskites

*Akash Dasgupta[1,2], Shuaifeng Hu[2], Seongrok Seo[2], Qimu Yuan[2], Yorrick Boeije[3], Michael Johnston[2], Sam Stranks[3], Henry Snaith[2]\**

1. Department of Chemistry, University of Washington, Seattle, WA 98195, USA
2. Clarendon Laboratory, Department of Physics, University of Oxford, Parks Road, Oxford OX1 3PU, UK.
3. Cavendish Laboratory, Department of Physics, University of Cambridge, JJ Thomson Avenue, Cambridge CB3 0HE, UK

## Abstract

Accurate optical constants are essential for modelling light propagation, absorption, and ultimately photovoltaic performance in state-of-the-art perovskite solar cells and is especially important for multiple junction or tandem cells. However, available datasets for metal-halide perovskites remain sparse, inconsistent in quality, and often suffer from unphysical sub-bandgap extinction caused by surface roughness and limitations of conventional ellipsometry fits. Here, we present a comprehensive library of complex refractive indices *(n,k)* for a technologically relevant set of FA-based lead perovskites, spanning bromide compositions from 0–100%, and mixed Pb–Sn perovskites with Sn fractions from 0–60%. Using state-of-the-art fabrication protocols that yield high-quality films, we combine variable angle spectroscopic ellipsometry measurements with highly sensitive sub-bandgap probes, including photothermal deflection spectroscopy for neat-lead based perovskites and Fourier-transform photocurrent spectroscopy for Pb–Sn alloys, to reconstruct "fully zeroed" dielectric functions across and below the band edge. The measured data are then stitched and recalculated via a Kramers–Kronig-consistent framework, ensuring physically accurate behaviour across the full spectral range. Finally, we introduce a transformation-based interpolation scheme that preserves spectral shape and feature alignment, enabling reliable determination of *(n,k)* for any intermediate composition or band gap. This complete dataset and interpolation protocol provide a standardized foundation for optical modelling of perovskite and tandem solar cells, addressing longstanding data gaps and supporting accurate simulations of next-generation photovoltaic architectures.

## 1 Introduction

The field of metal halide perovskites have had a veritable 'breakthrough decade'. Initially emerging as a promising new material for solar application, they showed modest PCEs with non-optimised fabrication and device architactures,[1,2] but exploded in interest once solid-state and thin film architectures had been demonstrated[3–5]. Since then, perovskites, when implemented in so called tandem architectures, have quickly risen to both the upper-end of efficiency charts (surpassing 34.6% for perovskite-silicon tandem)[6], as well as an integral technology likely to play a key role in meeting increasing societal energy needs[7]. In today's world, commercial sales of perovskite based solar technology to large industrial energy providers are underway[8] and likely to increase, proving definitively the real-world value of this technology.

In the early research into PSC, an area of interest was quantifying the upper limits of potential performance for PSCs[9–12]. To generalise, these works used measured absorption coefficients and/or 'external quantum efficiencies' (EQEs) from real materials or devices, to estimate maximum values for current-densities and voltages, by calculating the fraction of light absorbed by a device from the 'AM1.5' solar spectrum and the room temperature thermal black body, respectively. Evidently, obtaining good quality optical data forms the basis for this type of analysis.

As we move from single junction devices to multijunction tandems, the role of light management becomes increasingly important. In tandems, light passing through the device encounters many layers, each of different thicknesses and different optical densities, making reflection and interference effects significant. In the 'detailed balance' limit, where we assume (amongst other things) that the material absorption of each sub cell is unity above bandgap, Perovskite-on-Si tandems were predicted to have a maximum PCE of 45.1%[13], while studies which incorporated optical modelling predicted far more modest PCE limits of 32-34%[14,15]. Today, optical modelling continues to be employed to guide assessments, optimisations and discussions around perovskite-based tandem devices [16–19].

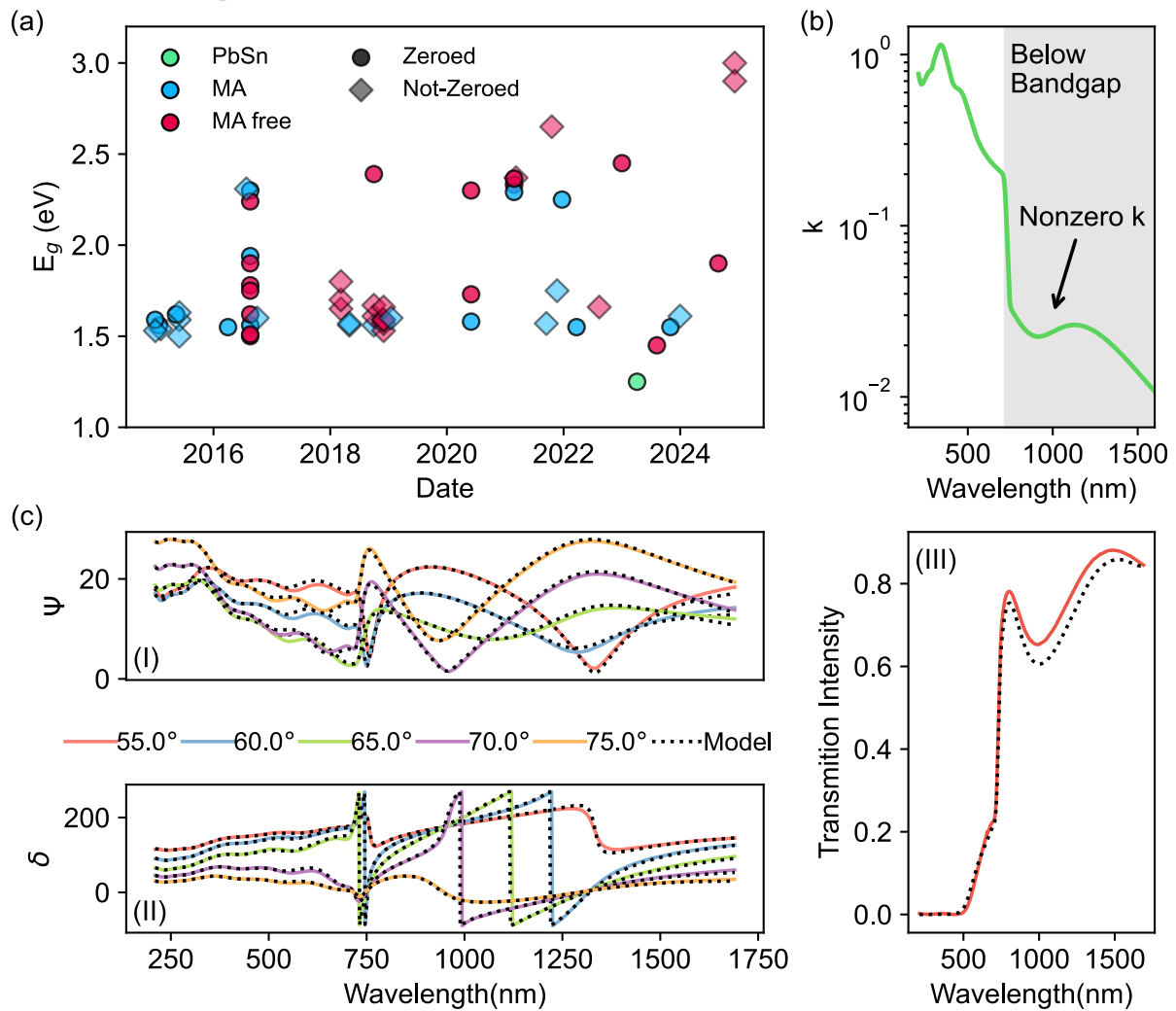

Figure 1: **(a)** Scatter plot depicting ellipsometry studies conducted on metal halide perovskites[17,20–47].. The plot shows publication dates versus the bandgap of the materials studied. Marker colours represent the materials: magenta markers indicate neat lead, methylammonium-free (MA-free) metal-halide perovskites, blue markers represent neat lead perovskites containing MA, and green markers represent mixed lead-tin metal halide perovskites. Diamond markers denote data where the reported extinction coefficient (k) does not approach zero past the band edge. **(b)** Example of a B-spline fit for the extinction coefficient (k) of a $Cs_{0.13}FA_{0.87}Pb(I_{0.75}Br_{0.26})_3$ perovskite. The logarithmic scale clearly shows that the k value does not approach

zero below the bandgap. **(c)** Variable angle spectroscopic ellipsometry measured psi (I) and delta (II), along with transmission data (III), for the fit shown in (b). The model fit is in good agreement with the data, despite the k value not reaching zero past the band edge.

Optical modelling (both in 'Transfer matrix' calculations as well as in 'Ray tracing' solvers) require the full complex refractive index of the material. That is, optical models require for each the values for $\tilde{n}(\lambda)$, where

$$\tilde{n}(\lambda) = n(\lambda) + i\kappa(\lambda), \qquad (1)$$

where $n$ is the refractive index, $\kappa$ the 'extinction coefficient', and $\lambda$ the wavelength. Both $n$ and $\kappa$ are required to describe the absorption, reflection and transition of light at interfaces. These values are commonly derived through 'variable angle spectroscopic ellipsometry' (VASE) measurements. However, the availability of good quality $\tilde{n}(\lambda)$ for a large range of perovskite composition remains a blind spot in the literature, which we seek to address here. In Figure 1a, we collect a range of $\tilde{n}(\lambda)$ values from literature reports across the years to illustrate the problem we identified. We propose that, to carry out optical simulations relevant to modern state-of-the art perovskite solar cells, there must be available $\tilde{n}(\lambda)$ values which meet the following criteria:

1. **Be contemporary:** The fabrication methods and compositions of metal-halide perovskites has evolved significantly in the last decade, leading to films with much stronger optical absorptions and narrower excitonic absorption peaks, a consequence of lower electronic disorder and less physical defects[48]. As a result, we require optical constants which reflect this improvement in material quality

2. **Be available across a relevant range of composition:** One of the powerful properties of perovskites are the tunability of the bandgap by controlling the stoichiometry. Hence, in order to explore or optimise, for example, a range of bandgaps for tandem applications, one requires $\tilde{n}(\lambda)$ values for a range of compositions with fine granularity. For Pb based perovskites there is a push to move away from using methylammonium as the A-site cation in the perovskite structure, opting for formamidinium for it's superior thermal stability [49–51], and incorporation of Sn into perovskites yields materials of lower (down to 1.23 eV) PV bandgap. Hence for modern simulations it's important to have FA based Pb perovskites and mixed PbSn containing perovskites represented in the optical constants data set.

3. **Be correctly "zeroed":** Perovskite thin films show a high surface roughness, which makes it difficult to accurately fit the $\tilde{n}(\lambda)$ values form the VASE measurement. This roughness can manifests as an unphysical absorption in the sub-bandgap region[52]. We demonstrate this in Figure 1b the fit $k(\lambda)$ for a ~1.8eV bandgap, as measured in our lab, shows apparent absorption at wavelengths within the bandgap, despite the 'fit' of the raw VASE data being relatively good (Figure 1c). While this is ~1-2 orders of magnitude under the above bandgap values, in calculations such as determining the dark saturation current, this low-level absorption will cause significant errors. Usable $\tilde{n}(\lambda)$ should be correctly zeroed, where the $k(\lambda)$ values appropriately fall to 0 past the band edge. Furthermore, for Sn-containing perovskites, there is often real sub-gap absorption due to the high levels of free-carriers as a consequence of Sn-vacancies. This creates measurable absorptions in VASE into plasmon states, which will not contribute to photocurrent generation[53].

With these criteria in mind, we can see in Figure 1a, the gaps in the current literature. In the early days of the field, there were more reports of $\tilde{n}(\lambda)$ values as researchers worked to characterise the material. Notably,

Ndione in 2016 reported $\tilde{n}(\lambda)$ values for a range of perovskite compositions, including FA based perovskites and the ellipsometry values that appear to be zeroed[44]. Werner and Subedi's efforts in 2018[24,47] taken together can span a range of 1.5-2.2eV using MA free perovskite, although these values were not zeroed. In any case, all will suffer from the limitations of fabrication methods available at the time, yielding $\tilde{n}(\lambda)$ values which are less representative for the state of the field today. Over time, we can recognise that when $\tilde{n}(\lambda)$ values are reported, they have generally been around 1.5-1.55eV, corresponding to a MA/FA/Cs-PbI$_3$ perovskite compositions, owing to the simplicity of the material as a basis system for study, as well as >2eV, representing generally novel wide bandgap materials being fabricated for the first time. Mixed Pb/Sn perovskites are badly represented in the literature, with only a handful of reports.

In this work we address these shortcomings, by providing a complete series of $\tilde{n}(\lambda)$ values which can form a standard library for the community to draw upon for the purposes of modelling and simulations. By taking VASE measurements for a range of FA/Pb based perovskites at varying Br content and mixed PbSn perovskites at a range of Sn compositions, we provide optical constants for the range of bandgaps and compositions relevant for tandem solar cells[14]. We use state-of-the-art fabrication processes which have been used in our research group to produce record efficiency devices[54], ensuring the quality of our optical constants are representative of the advances made in fabrication processes in the last decade. To overcome the zeroing issue that was present in many of our samples, instead of using different fitting and corrections as others have done in the past, we directly use additional techniques to accurately measure the optical constants near and past the band edge, including photothermal deflection spectroscopy and Fourier-transform photocurrent spectroscopy. Finally, in order to provide an optical constant for any arbitrary composition, we also devise an interpolation scheme, which allows us to obtain an estimated optical constant for stoichiometry between those that we explicitly measured as part of this work. We hope that this scheme will prove to be a valuable addition to the literature and provide a useful resource for future researchers to obtain the optical constants for their simulation work[55].

## 2  Ellipsometry measurement

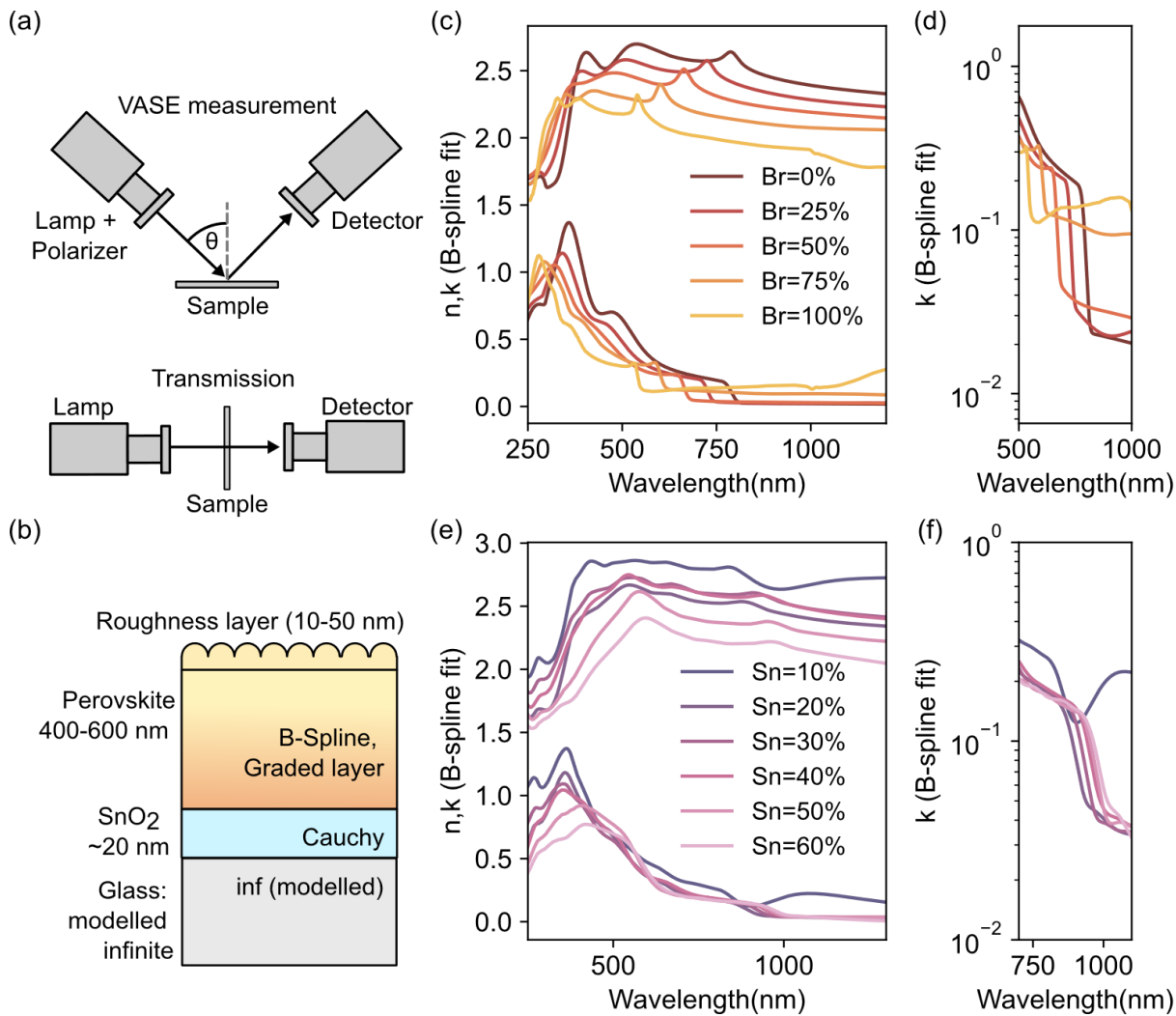

Figure 2: **(a)** Schematic of the measurement setup used to obtain data for fitting optical constants, including variable angle spectroscopic ellipsometry (VASE) and transmission measurements, both performed with the JA Wollam RC2 ellipsometer. **(b)** Model used to fit the ellipsometry data. The underlying layers (Glass, Tin Oxide) were measured independently. The perovskite layer was modelled as a graded layer using a B-spline approach, including surface roughness, with the JA Wollam Complete Ease software. The fit was not constrained to zero beyond the band edge. **(c,d)** Refractive index (n) and extinction coefficient (k) B-spline fit of the ellipsometry data for $Cs_{0.13}FA_{0.87}Pb(I_{1-x}Br_x)_3$ perovskites with varying bromide content (x=0,0.25,0.5,0.75,1). The log plot shows that k does not approach zero past the band edge. **(e,f)** Refractive index (n) and extinction coefficient (k) B-spline fit of the ellipsometry data for $Cs_{0.1}FA_{0.61}MA_{0.3}Sn_xPb_{(1-x)}I_3$ perovskites with varying tin content (x=0,0.2,0.3,0.4,0.5,0.6). The log plot demonstrates that k does not approach zero past the band edge.

To begin with, we will first examine the ellipsometry measurements done in their raw form without additional corrections. For this work, we measured a range of compositions as mentioned in the previous section. Specifically, we measured $Cs_{0.13}FA_{0.87}Pb(I_{1-x}Br_x)_3$ (x=0,0.25,0.5,0.75,1) and $Cs_{0.1}FA_{0.61}MA_{0.3}Sn_xPb_{(1-}$ $_x$)I$_3$ (x=0,0.2,0.3,0.4,0.5,0.6). The former represents a range of material with band gaps spanning between 1.5 to 2.2 eV while the latter represents lower band gaps down to around 1.2 eV[50]. Full details can be found in the method section, but briefly, for each sample measured we carried out both a VASE measurement, as well as a simple transmission measurement done on the same ellipsometer (Figure 2a). Having the additional transmission data incorporated into the fits provided an additional constraint which insured our fit were as physical as possible. This data was fit to a model which considered an infinite glass substrate, a tin-oxide adhesion layer, and a roughness layer at the perovskite/air interface, as we have illustrated in (Figure 2b). The perovskite fits were done using a B-Spline model. A simple grading scheme was used to account for variations in the dielectric function caused by air degradation, which we assumed would act on the top of the film more than on the bottom.

Looking at the fits obtained, we observed that as we increase the bromide content, the energies at which the peaks appear in the $k(\lambda)$ values shift towards higher energies, with the relative height and shapes changing slightly (Figure 2b). Near the band edge, the excitonic nature of the peak manifest itself more as we move to higher bromide content. It is clear, however, that we are not able to represent the band-edge accurately from this measurement alone, with the unphysical absorption past the band-edge being present in all the measurements and becoming more pronounced for higher bromide content (Figure 2c). For the mixed lead-tin samples, the change in the optical constants have a slightly different trend. In this case, instead of the features shifting along the wavelength axis, we observe a high energy peak (~375nm) being supressed while a different, broader peak (~420nm) is expressed (Figure 2e). This is likely due to the change in the density of states of the material due to the interactions between the orbitals of the lead and the tin ions, but the specifics of this is beyond the scope of this paper. The band edge fits on these samples are worse than the neat lead samples, with the difference in band gap between the different tin compositions being barely resolvable (Figure 2f). This is likely due to both the greater surface roughness of the material, as well as the significant self p-doping in mixed lead-tin perovskites[56,57], which leads to free carrier absorption in the below-bandgap region[53].

The non-physical band edge is problematic when calculating quantities such as the expected equilibrium carrier concentration in the material for solar cell applications, in addition to maximum photocurrent generation in an absorber layer. In fact, the shape of the band edge many orders of magnitude below the onset can significantly influence the dark recombination current density calculations. Therefore, in our effort to produce a set of high-quality, "properly zeroed" dielectric constants, we address this issue. Enforcing a simple transparent region by forcing the k values to go to zero beyond a specified wavelength led to poor fits. We found that the band-edge shape could change significantly and sometimes take on an unnatural form depending on the arbitrary choice of the cutoff wavelength. We also attempted to account for surface roughness by adding an effective-medium layer to the fitted stack, but this approach did not allow us to fit the bandgap or sub-bandgap region effectively. Additionally, for the mixed lead–tin samples, a Drude-like absorption could be used to fit parts of the data; however, because this absorption overlaps with the true band edge, it becomes difficult to disentangle free-carrier absorption from band-to-band absorption. In such cases, the VASE measurement alone may not contain sufficient information to accurately reproduce the band-edge shape. For these reasons, we decided to augment the fitted dielectric constants with additional measurements and stitch the data together to produce the most accurate dataset possible.

# 3 Stitched band edge

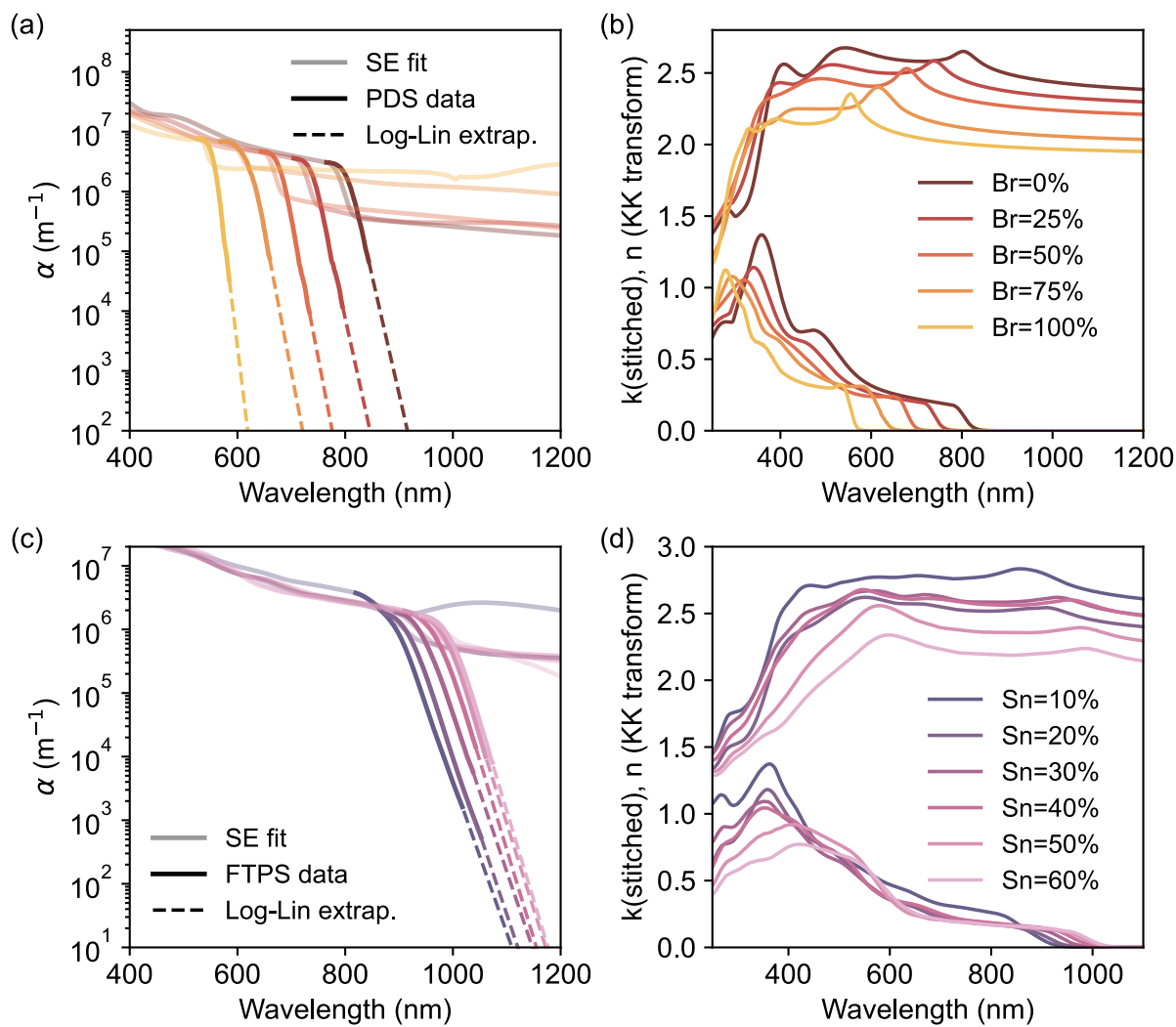

Figure 3: (a) Absorption coefficient from the B-spline fit of ellipsometry data for $Cs_{0.13}FA_{0.87}Pb(I_{1-x}Br_x)_3$ (x=0,0.25,0.5,0.75,1) perovskites, with Photothermal Deflection Spectroscopy (PDS) data overlaid. The PDS data is scaled to match the B-spline fit at the stitch wavelength near the band edge. A log-linear extrapolation of the band edge beyond the PDS noise floor is included in the final stitched plot. Bromide concentration is indicated in (b). (b) Processed optical constants for $Cs_{0.13}FA_{0.87}Pb(I_{1-x}Br_x)_3$ (x=0,0.25,0.5,0.75,1) perovskite compositions. The extinction coefficient (k) combines the spectroscopic ellipsometry fit, PDS data, and the extrapolated band edge from (a). The refractive index (n) is recalculated using the Kramers-Kronig transform, anchored by the SE-derived nnn value near the band edge. (c) Absorption coefficient from the B-spline fit of ellipsometry data for $Cs_{0.1}FA_{0.61}MA_{0.3}Sn_xPb_{(1-x)}I_3$ (x=0,0.2,0.3,0.4,0.5,0.6) perovskites, with Fourier Transform Photocurrent Spectroscopy (FTPS) data overlaid. The FTPS data is scaled to match the B-spline fit at the stitch wavelength near the band edge, assuming the EQE absorption onset aligns with the non-free carrier absorption onset. A log-linear extrapolation of the band edge beyond the FTPS noise floor is included in the final stitched plot. Tin concentration is indicated in (d). (d) Processed optical constants for $Cs_{0.1}FA_{0.61}MA_{0.3}Sn_xPb_{(1-x)}I_3$ (x=0,0.2,0.3,0.4,0.5,0.6) perovskite compositions. The extinction coefficient (k) combines the spectroscopic ellipsometry fit, FTPS data, and the extrapolated band edge from (c). The refractive index (n) is recalculated using the Kramers-Kronig transform, anchored by the SE-derived n value near the band edge.

To augment our VASE measurements, we employed additional high-sensitivity techniques to obtain a more accurate determination of the imaginary component of the dielectric function near the absorption edge. For the neat lead materials, we used photothermal deflection spectroscopy (PDS) on samples deposited directly onto glass substrates. PDS is a highly sensitive technique that detects changes in the local refractive index of the surrounding medium induced by heat generated from optical absorption (see Methods for full details). For our samples, this technique provided sensitivity extending roughly three orders of magnitude below the absorption onset and was unaffected by the surface roughness that limited the VASE measurements.

PDS yields absorption coefficient ($\alpha$) in arbitrary units, which is related to $\kappa$ through the simple relationship:

$$\alpha = \frac{4\pi\kappa}{\lambda}, \quad (2)$$

where $\lambda$ is the wavelength of excitation. The VASE measurements already provide $\kappa(\lambda)$ with reliable fits up to a point near the band edge, beyond which the fits become unphysical. To address this, we "stitched'' the PDS-derived arbitrarily scaled absorption spectra to the VASE-derived $\kappa(\lambda)$ by replacing the unreliable portion of the VASE fit with the corresponding PDS data. Because the PDS signal is in arbitrary units, we normalised it to the VASE values at the stitch point, ensuring smooth continuity between the two datasets. The resulting stitched data is illustrated in Figure 3a, where the VASE data is rendered in transparent colour, while the PDS measured data (after normalisation) is depicted with in solid colour and overlaid. We plot the PDS data up to the chosen switch-over point. Since the PDS-derived band-edge shape is considerably more reliable than the VASE fits, we selected this switch-over point to span the entire band-edge region, including the "turnover" indicative of the transition from below to above band gap energy, and a small extension into the higher-wavelength regime. Finally, to capture the exponentially decaying tail of the band edge, which continues over many orders of magnitude below the PDS detection limit, we performed a log-linear fit to the band-edge region and extrapolated beyond the measurable range. This extrapolated tail is also shown in Figure 3a with dotted lines.

The final stitched values for $\kappa(\lambda)$ for the different neat-lead compositions are shown in Figure 3b. In contrast to the plots in Figure 2c, the stitched data now correctly approach zero at sub-bandgap wavelengths while retaining the same overall behaviour in the higher-wavelength region. The real component of the complex refractive index, $n(\lambda)$, is not independent of $\kappa(\lambda)$ and must be recalculated to be consistent with the new stitched $\kappa(\lambda)$. This can be calculated using the Kramers-Kronig relationship:

$$n(\omega) = n_\infty + \frac{2}{\pi} \int_0^\infty \frac{\Omega \kappa(\Omega)}{\omega^2 - \Omega^2}, \quad (3)$$

where $\omega = 2\pi/\lambda$, and $n_\infty$ is the refractive index at infinite frequency. We compute this quantity numerically by modifying the so-called "Singly Subtractive Kramers-Kronig Relation" method[58,59], which requires specification of an "anchor" frequency, $\omega_a$, at which the refractive index $n_a$ is known. We determine $n_a$ by referencing the refractive index obtained from the VASE fits. Specifically, by evaluating the integral term in Eq. (3) at the stitch point using the newly stitched $\kappa(\lambda)$, the anchor refractive index follows as the difference between this integral and the VASE-determined refractive index:

$$n_a = n_{VASE}(\omega_{st}) - \frac{2\omega_{st}}{\pi} \int_0^\infty \frac{\Omega \kappa(\Omega)}{\omega_{st}^2 - \Omega^2}, \quad (4)$$

where $\omega_{st}$ is the angular frequency corresponding to the stitch point. Comparing the recalculated values of $n(\lambda)$ in Figure 3b to those obtained from the ellipsometry fits directly in Figure 2c, we find good agreement in the above-bandgap region, extending down to approximately 300 nm. For shorter wavelengths than 300 nm, the values derived from the ellipsometry fit and those obtained from the Kramers–Kronig recalculation begin to diverge. This divergence arises because the measured data are truncated at 200 nm, leading to a poor approximation of the integral in this region, which in principle should extend to infinite energy. The proprietary 'CompleteEASE' software[60] used for the VASE fitting accounts for this limitation through several correction strategies; however, we did not incorporate these adjustments into our model. Wavelengths below 300 nm are generally less relevant for solar applications, since both the solar spectrum and the thermal black-body spectrum at room temperature have minimal intensity in this range. Additionally, these short wavelengths are typically subject to parasitic absorption or intentional filtering elsewhere in a fully fabricated device or module. For these reasons, we report optical constants only down to 300 nm, corresponding to the spectral regime in which we maintain high confidence in the calculated values.

For the mixed lead–tin materials, we carried out a similar procedure; however, instead of using PDS, we employed Fourier-transform photoexcitation spectroscopy (FTPS) in fabricated solar cells, in order to accurately determine the absorption at and within the band gap of the lead-tin perovskite absorber layers. FTPS requires fabrication of a fully operational device stack, since it relies on electrically contacting the sample. Consequently, the measured band-edge response is inevitably convolved with the absorption, transmission, reflectance and interference characteristics of the additional layers in the device structure. However, as noted previously, the mixed lead–tin materials exhibit significant free-carrier absorption: light is absorbed by equilibrium carriers rather than generating new photoexcited carriers. This absorption is prominent in the sub-band-gap region and does not contribute to photocurrent generation or recombination in the device. Therefore, it should not be included when calculating quantities such as the dark saturation current or the optical current densities under illumination in a solar cell. Although free-carrier absorption can act as a form of parasitic loss, this is generally not relevant in practical tandem architectures because the lead–tin perovskite is typically the bottom cell, possessing the narrowest bandgap. In terms of competition between above band gap absorption in a perovskite absorber, with the free-carrier absorption, attenuation in the latter is <<1%. For these reasons, despite the added uncertainty of contribution to the band edge shape from the optical response of the full device stack, FTPS was chosen for this measurement; It selectively probes photo-generated charge carriers and is not affected by free-carrier absorption, aside from the fractional attenuation due to parasitic free-carrier absorption. Full experimental details for this method are provided in the Methods section.

Apart from the different methods used to probe the band edge, the remainder of the analysis for the mixed lead–tin samples follow the same procedure as for the neat-lead samples. Specifically, we stitched the band-edge absorption measurement to the ellipsometry-derived $\kappa$ spectrum at a point slightly above the band-edge onset and applied a normalization to ensure smooth continuity. We then recalculated $n(\lambda)$ and $\kappa(\lambda)$ using the Kramers–Kronig transformation as described above. The resulting optical constants are shown in Figure 3 c,d.

# 4 Interpolation scheme

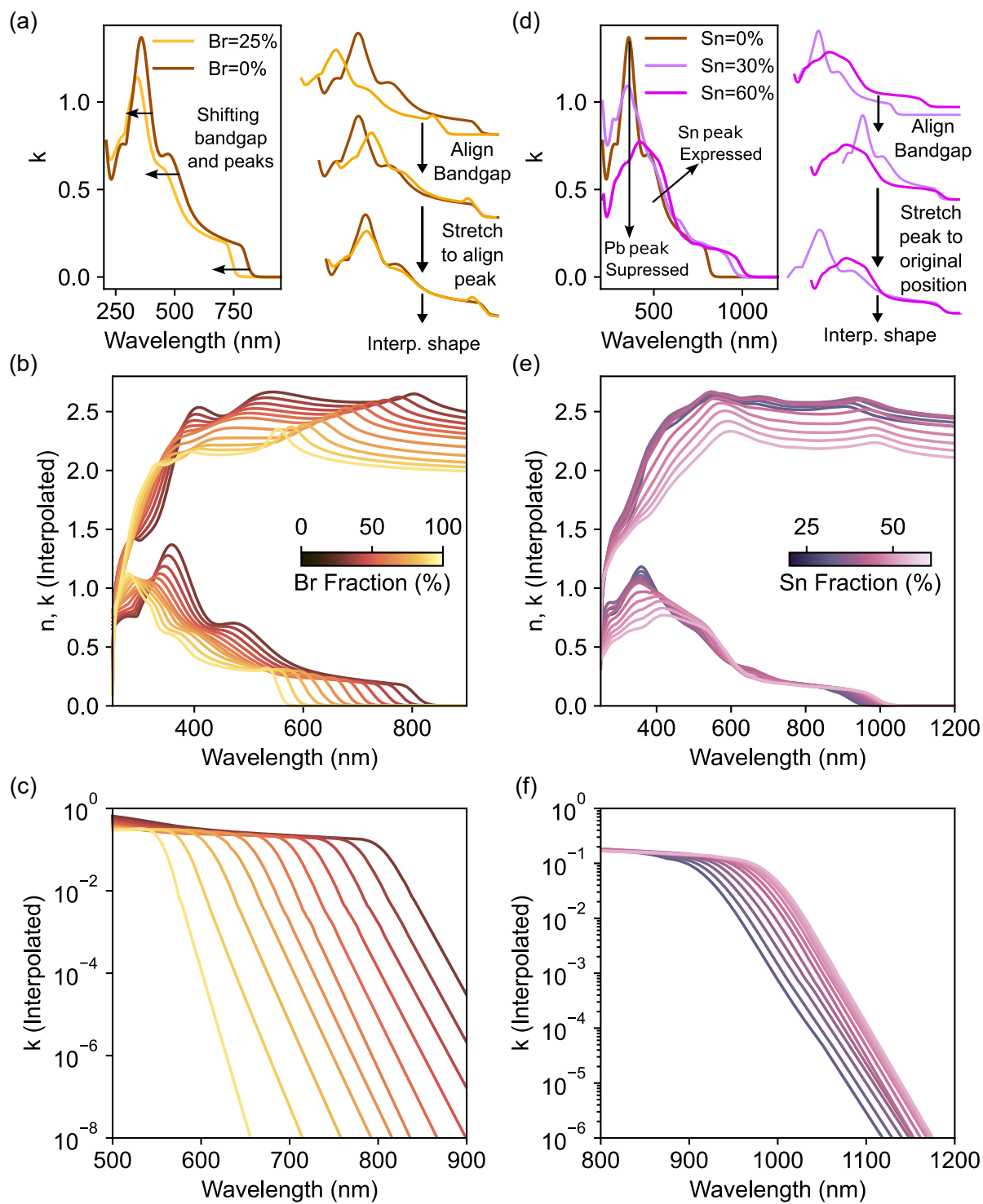

Figure 4: **(a)** Illustration of how the extinction coefficient changes for $Cs_{0.13}FA_{0.87}Pb(I_{1-x}Br_x)_3$ perovskites as the bromide content is changed, and the transformations applied to the curves for shape interpolation. Because all the peaks shift, the curves were shifted and stretched such that the band edge and peaks line up **(b)** Interpolated refractive index (n) and extinction coefficient (k) for $Cs_{0.13}FA_{0.87}Pb(I_{1-x}Br_x)_3$ perovskites, at gradually varying bromide fraction. The values depicted are interpolated from the data in Figure 3b. **(c)** Log plot of the interpolated extinction coefficient (k) for $Cs_{0.13}FA_{0.87}Pb(I_{1-x}Br_x)_3$ perovskites, illustrating the band edge more clearly. **(d)** Illustration of how the extinction coefficient changes for $Cs_{0.1}FA_{0.61}MA_{0.3}Sn_xPb_{(1-x)}I_3$ perovskites as the tin content

is changed, and the transformations applied to the curves for shape interpolation. Here, one peak is supressed while a different peak is expressed, so the curves were shifted and stretched such that the band edge lines up, but the peaks are returned to the original energies **(e)** Interpolated refractive index (n) and extinction coefficient (k) for $Cs_{0.1}FA_{0.61}MA_{0.3}Sn_xPb_{(1-x)}I_3$ perovskites, at gradually varying tin fraction. The values depicted are interpolated from the data in Figure 3b. **(f)** Log plot of the interpolated extinction coefficient (k) for $Cs_{0.1}FA_{0.61}MA_{0.3}Sn_xPb_{(1-x)}I_3$ perovskites, illustrating the band gap change more clearly.

Finally, to facilitate carrying out optical simulations for an arbitrary bandgap/material stoichiometry, we need to be able to effectively interpolate the dielectric functions for an arbitrary material composition based on the experimentally measured compositions described thus far. Direct interpolation of $\kappa(\lambda)$ at each wavelength introduces substantial artefacts, because the characteristic features such as the absorption onset and higher-energy critical points shift with composition. Hence, we require an interpolation scheme which effectively captures the changing shape of the functions, as well as the relative shift along the wavelength axis. To this end, we developed a transformation-based interpolation protocol that preserves the physical shape and spectral alignment of these features.

To begin, we will consider the neat lead materials. For an intermediate bromide fraction $x$ (such that the targeted composition is $Cs_{0.13}FA_{0.87}Pb(I_{1-x}Br_x)_3$), we can first identify two values of $\kappa(\lambda)$ from our measured set: $\kappa_1$ and $\kappa_2$, for respective bromide fractions of $x_1$ and $x_2$ such that $x_1 \leq x \leq x_2$. We can estimate the bandgap energies for these materials ($E_g^1, E_g^2$ respectively) through the Tauc-plot method[61]. We note that although the Tauc band gap is not physically meaningful for a material with strong excitonic absorption at the band edge[62], nor is it the important band gap to consider for PV devices, where the "PV-band gap" is often taken from the inflection point of the EQE[63], it *is* a property of the absorber material. It can therefore be used to quantify and catalogue different perovskite optical band gaps independent of how the material is integrated into a solar cell or other optoelectronic device[64]. Each function is then transformed to a common reference frame through two steps:

1. **Band-edge alignment:** Both spectra are shifted along the wavelength axis so that their band edges coincide at zero nm.

$$\kappa_1'(\lambda) = \kappa_1\left(\lambda + \frac{hc}{E_g^1}\right), \quad \kappa_2'(\lambda) = \kappa_2\left(\lambda + \frac{hc}{E_g^2}\right) \quad (5)$$

2. **High-energy peak alignment:** The higher-energy critical points of $\kappa_2$ are aligned to those of $\kappa_1$ by applying a wavelength-axis stretching factor, ensuring that the spectral features overlap in position as well as shape

$$\kappa_2'' = \kappa_2'\left(\frac{\lambda}{\sigma_{1,2}}\right), \quad (6)$$

where $\sigma_{1,2}$ is the stretching factor required to align the critical points.

This process is illustrated in Figure 4a. Once aligned, a linear interpolation between the transformed spectra yields a provisional $\kappa$ spectrum for the desired bandgap:

$$\kappa_x'' = \kappa_1' + \left(E_G^x - E_g^1\right) \cdot \frac{\kappa_2'' - \kappa_1'}{E_g^2 - E_g^1}, \quad (7)$$

This intermediate spectrum is then "un-stretched" using an interpolated stretching factor and translated back to its physical targeted wavelength position (or Tauc energy gap)

$$\kappa_x = \kappa_x'' \left( \lambda \cdot \sigma_i - \frac{hc}{E_g^x} \right), \sigma_i = \sigma_{1,2} \cdot \frac{E_g^x - E_g^1}{E_g^2 - E_g^1} \tag{8}$$

restoring the correct spectral shape and band-edge location. This procedure preserves both the relative feature positions and the characteristic line shapes, avoiding the distortions that arise from direct point-wise interpolation. Finally, the interpolated $n_x(\lambda)$ can be calculated by doing the Kramers-Kronig transform. The anchor refractive index is also interpolated in the same way as the $\kappa_x$ values, that is, applying the transformations in Eq. (5 and (6 to the corresponding $n_1$ and $n_2$, and interpolating the value of the anchor points near the band edge onset.

The resulting interpolated dielectric functions are shown in Figure 4b, with the band edge displayed more clearly on a logarithmic scale in Figure 4c. We observe that, using this scheme, the interpolated values shift smoothly across the wavelength axis while preserving the relative spectral shape. In addition, the gradients of the band edge in the logarithmic representation evolve appropriately, indicating that the interpolation captures the expected physical trends.

In the case of the mixed lead–tin materials, we applied a similar interpolation scheme; however, the transformations required were necessarily different. This difference arises because, as seen when comparing Figure 4a and Figure 4e, the evolution of κ as a function of stoichiometry is qualitatively distinct from that of the neat-lead series. For the neat-lead samples, both the band-edge onset and the higher-energy critical points shift coherently along the wavelength axis. This behaviour is consistent with prior DFT studies, which show that changes in halide composition lead primarily to uniform shifts in the band structure[65]. In contrast, for the mixed lead–tin materials, only the band edge shifts in a straightforward manner. The higher-energy critical points do not translate along the wavelength axis; instead, they evolve in a qualitatively different way. As the tin content increases, one spectral feature (at ~375 nm) is gradually suppressed while another (at ~420 nm) becomes more prominent. This behaviour suggests a fundamentally different modification of the electronic structure in the tin-containing alloys, although a full theoretical investigation lies beyond the scope of this work.

For our purposes, it is sufficient to recognise this qualitative difference and incorporate it into the design of our interpolation scheme. As before, we first transform each κ spectrum so that the band edges align, as described in Eq. (5. However, for the mixed alloys, the stretching factor in Eq. (6 is defined differently. The relevant higher-energy feature appears near 375 nm for the 0% Sn composition. Rather than stretching the transformed κ so that the critical points align directly, we instead stretch the spectra so that the region originally located at 375 nm (before translation) coincides across compositions. The remainder of the interpolation procedure is identical to that used for the neat-lead case, with the results illustrated in Figure 4f and Figure 4g.

# 5 Reliability and limitations

We will take a moment here to discuss the reliability of the data set presented in this work. The dielectric functions shown in Figures 3b and 3d are reconstructed quantities, each derived by stitching a sub-bandgap absorption measurement to ellipsometry data, followed by Kramers–Kronig recalculation of $n(\lambda)$. As such, they represent well-justified approximations rather than direct measurements.

For the neat-lead materials, the PDS-based stitching procedure yields a $\kappa(\lambda)$ spectrum that we consider highly reliable. The method is straightforward, requires minimal assumptions, and has been extensively validated in the literature. The corresponding Kramers–Kronig calculation is also robust within the wavelength range where the measured dataset is complete. As noted previously, however, truncation of the ellipsometry data at short wavelengths limits the validity of the recalculated $n(\lambda)$ to approximately 300 nm; beyond this point, the integral no longer accurately reflects the true higher-energy contributions.

For the mixed lead–tin compositions, the situation is inherently more complex. The FTPS measurement provides an effective external quantum efficiency (EQE), and our procedure assumes that this signal is proportional to the material's absorption coefficient:

$$EQE(\lambda) \propto \alpha(\lambda). \tag{9}$$

We consider this assumption to be reasonable. Although parasitic absorption or reflection from other layers may contribute to the shape of the measured EQE, normalizing the EQE to the values obtained from the VASE measurement compensates for any reduction in absolute magnitude. Furthermore, the spectral absorption of the additional layers in the device stack is expected to follow either a Cauchy-like or Drude-like behaviour in the infrared, both of which vary only slowly with wavelength. This justifies the assumption that the rapid changes in the measured slope are dominated by the optical response of the active layer itself. Nevertheless, we acknowledge that this proportionality introduces a possible source of systematic error. We highlight again however that it remains preferable to the alternative: relying on ellipsometry alone or using the PDS measurement as in the case of the neat lead samples, where free-carrier absorption would obscure the true band edge entirely.

Our interpolation scheme itself is semi-qualitative by construction. By aligning and stretching spectra to match key spectral features, we preserve their physical behaviour but do not claim a fully ab initio treatment. As a result, validation is necessarily limited to qualitative consistency with the measured datasets at the discrete compositions. This limitation does not compromise the reliability of the measured optical constants; rather, it affects only the interpolated values used to generate continuous trends across bandgap space. To maintain transparency and ensure that the community may adopt whichever dataset best suits their needs, we will release both the optical constants derived from our experimentally fabricated films, and the interpolated dielectric functions. Users may therefore choose between the directly measured spectra and the interpolated set depending on the requirements of their modelling or device-simulation work[55].

# 6 Conclusion

In this work, we address a critical limitation in the perovskite optoelectronics community: the lack of reliable, compositionally or band gap comprehensive optical constants for modern high-quality metal-halide perovskite films. By combining advanced VASE analysis with high-sensitivity band-edge measurements (PDS for neat-lead compositions and FTPS for mixed Pb–Sn alloys), we overcome the unphysical sub-bandgap extinction commonly encountered in ellipsometry-only datasets. The resulting stitched and Kramers–Kronig-recalculated dielectric functions exhibit physically accurate band-edge behaviour, while retaining fidelity to the above-bandgap spectral features relevant for optical modelling.

Beyond providing high-quality reconstructed $(n, \kappa)$ datasets at discrete compositions, we also introduce a transformation-guided interpolation strategy that accounts for the distinct ways bromide incorporation and tin alloying modify spectral features. This enables smooth, physically meaningful interpolation across composition space, an essential capability for designing tandem architectures and exploring device optimisation through simulation.

Together, the datasets (with 10meV granularity for the interpolated set) and interpolation framework constitute a unified optical-constant library tailored to the needs of modern perovskite research. By releasing both raw stitched measurements and interpolated functions, we aim to provide a transparent foundation upon which the community can build accurate optical models, evaluate theoretical efficiency limits, and guide the design of perovskite/silicon, all-perovskite and other hybrid perovskite-based tandem devices. This work thus bridges a longstanding gap in optical data availability and enables more robust modelling of next-generation photovoltaic technologies.

# 7 Methods

## 7.1 Sample Preparation

### 7.1.1 Precursor material preparation for pure-Pb perovskites

Lead iodide (PbI2, 99.99%), lead bromide (PbBr2, > 98.0%), Caesium iodide (CsI, > 99.0%), and Caesium bromide (CsBr, > 99.0%) and were purchased from TCI. Formamidinium iodide (FAI, > 99.99%) was purchased from Dyenamo. Formamidinium bromide (FABr, 99.99%) was purchased from Sigma-Aldrich. Unless stated otherwise, all other materials and solvents were purchased from Sigma-Aldrich. To form 1.2M the double-cation lead triple-halide perovskite precursor solutions, CsI, CsBr, FAI, FABr, PbI2, and PbBr2 were weighed to obtain stoichiometric $Cs_{0.17}FA_{0.83}Pb(I_xBr_{1-x})_3$ hybrid perovskite composition. The powders were dissolved in N, N-dimethylformamide (DMF): dimethyl sulfoxide (DMSO) solution (4:1 volume) for I1Br0, I0.75Br0.25, and I0.5Br0.5 compositions, and 1:1 volume for I0.25Br0.75, and I0Br1 compositions due to solubility. The solutions were stirred 2-3 hours in a nitrogen-filled glovebox before use.

### 7.1.2 Film deposition for pure-Pb perovskites

Glass substrates were cleaned by sonicating substrates in various solutions in the following sequence: 1) deionized water with 2% v/v solution of Decon 90 cleaning detergent; 2) deionized water; 3) acetone and 4) isopropanol (each step for 15 minutes). After 30 minutes of UV-O3 treatment, 40nm ALD-Al2O3 process was deposited to increase wettability of substrate and reduce voids at substrate/perovskite interface. After 30 minutes of UV-O3 treatment, substrates were transferred to a nitrogen-filled glove box. the perovskite layer was formed by spin coating a 1.2M perovskite solution starting at 1000 rpm for 10 seconds (ramping time of 5 seconds from stationary status) and then 5000 rpm (ramping time of 4 seconds from 1000 rpm) for 35 seconds. Before the end of the spinning process, a solvent-quenching method was used by dropping Anisole (325 μL) onto the spinning substrates at 40 s after the start of the spin-coating process. The perovskite is annealed for a total of 30 minutes at 100° C.

### 7.1.3 Precursor material preparation Mixed Pb-Sn Materials

Unless otherwise stated, all materials were used as received without further purification. Methylammonium iodide (MAI, >99.0%), and formamidinium iodide (FAI, >98.0%), were purchased from Greatcell Solar Materials. Bathocuproine (BCP, >99.0%), and lead iodide ($PbI_2$, 99.99%, trace metals basis) were purchased from Tokyo Chemical Industry Co., Ltd. (TCI). Caesium iodide (CsI, 99.999%, metals basis) was purchased from Alfa Aesar. Ammonium thiocyanate (NH$_4$SCN, 99.99% trace metals basis), tin fluoride (SnF$_2$, 99%), tin iodide (SnI$_2$, beads, 99.99%, trace metals basis), and Sn(0) powder (<45 μm particle size, 99.8% trace metals basis). Poly(3,4-ethylenedioxythiophene):poly(styrene sulfonate) (PEDOT:PSS) aqueous solution (Clevios PVP AI 4083) was purchased from Heraeus Co., Ltd. Fullerene C$_{60}$ (sublimed, 99.99%) was purchased from CreaPhys GmbH. Dehydrated dimethylsulfoxide (DMSO), dehydrated *N,N*-dimethylformamide (DMF), and chlorobenzene (CB) were purchased from Sigma-Aldrich. Tetrakis(dimethylamino) tin(IV) was purchased from Pegasus Chemical Limited.

### 7.1.4 Film deposition for Mixed PbSn perovskite thin films

The perovskite film was prepared in an N$_2$-filled glove box (H$_2$O, O$_2$ < 0.1 ppm). The Cs$_{0.1}$FA$_{0.6}$MA$_{0.3}$Sn$_x$Pb$_{(1-x)}$I$_3$ (x = 0.1, 0.2, 0.3, 0.4, 0.5, and 0.6) perovskite precursor solutions were prepared by mixing designated amount of CsI, FAI, MAI, SnI$_2$, PbI$_2$, SnF$_2$ (10 mol% with respect to SnI$_2$), and NH$_4$SCN (4 mol% with respect to SnI$_2$) in a solvent mixture of DMSO and DMF (v/v, 1/3) to reach a concentration of 1.50 M for all the samples[66,67]. 0.5 mg mL$^{-1}$ Sn(0) powder was added to the perovskite precursor solution to scavenge the potential certain amount of Sn(IV)[68]. The precursor solutions were stirred at 45 °C for about 40 min and filtered through a 0.20 μm PTFE filter before use. To spin coat the films, 200 μL of the room temperature precursor solution was applied to the substrate. A two-step spin coating program was used. The first step was 1000 rpm for 10 s with an acceleration of 200 rpm s$^{-1}$, and the second was 4000 rpm for 40 s with a ramp-up of 1000 rpm s$^{-1}$. Room-temperature chlorobenzene (500 μL for each of the 30 mm by 30 mm substrates) was used as the antisolvent. The chlorobenzene was quickly dripped onto the surface of the spinning substrate over an interval of 1 s during the second spin coating step 20 seconds before the end of the procedure. The substrate was then immediately annealed on a 100 °C hot plate for 10 min, followed by annealing at 65 °C for over 10 min to avoid glovebox vapour ingress of the as-prepared films, then the films were cooled down to room temperature for the following processes[66,67]. The spin coating process was set as 4000 rpm for 20 s with an acceleration of 1333 rpm s$^{-1}$. Following spin coating, the films were immediately annealed again at 100 °C for about 5 min. For the films prepared for the ellipsometry measurements, a 20 nm ALD-SnO$_X$ layer was grown before the perovskite layer.

### 7.1.5 Fabrication of single-junction mixed tin–lead perovskite solar cells

The patterned glass/FTO substrates (15 Ω sq$^{-1}$, Latech Scientific Supply Pte. Ltd. or 10 Ω sq$^{-1}$, AGC Inc.) were consecutively cleaned with 15 min ultrasonic bath in water, detergent solution, acetone, and isopropanol, followed by drying with an N$_2$ gun, and finally plasma treatment right before the following deposition. The PEDOT:PSS hole transport layer was fabricated from an aqueous dispersion (diluted, PEDOT:PSS dispersion/water, 4/1, v/v), which was filtered through a 0.45 μm PVDF filter and then spin-coated on the FTO substrate using a spin program of 10 s at 500 rpm followed by 30 s at 5000 rpm. The films were then annealed in air at 140 °C for 20 min. After transferring to an N$_2$-filled glove box (H$_2$O, O$_2$ < 0.1 ppm), the substrates were degassed at 140 °C for about 10 min. The perovskite layer was fabricated on PEDOT:PSS following the above-mentioned procedure. The samples were moved under N$_2$ to a vacuum deposition chamber, where 20 nm of C$_{60}$ (deposition rate 0.01 nm s$^{-1}$) and 8 nm of BCP (deposition rate 0.01 nm s$^{-1}$) were deposited by thermal evaporation. The top electrode was prepared by depositing 100 nm of silver (Ag) through a shadow mask. The deposition rate for Ag was firstly set as 0.03 nm s$^{-1}$ to reach 10 nm, then raised to 0.07 nm s$^{-1}$ to reach 40 nm, and finally raised to 0.1 nm s$^{-1}$ to reach the target thickness. All cells were encapsulated with a cover glass and UV-activated adhesive (Eversolar AB-341, Everlight

Chemical Industrial Co.) which was cured under a UV-LED lamp (peak emission at 365 nm) for 3 min inside an $N_2$-filled glovebox ($H_2O$, $O_2$ < 0.1 ppm) before any related characterization under ambient conditions.

Unless otherwise stated, all the vacuum-based depositions were processed with the National Thin Film Cluster Facility for Advanced Functional Materials (NTCF) assembled by Angstrom Engineering Inc. The Picosun branded model R200 Advanced Plasma ALD chamber was connected and controlled by a main cluster operation system. The ALD system was vacuum-pumped by a two-stage Edwards iXH610 pumping system, and the process pressure was usually maintained at 5 mbar. The background carrier gas for the ALD system was $N_2$ (99.9999%). The ALD was equipped with two room-temperature sources for liquid-based precursors (typically de-ionised water and trimethylaluminum (TMA, Pegasus) were loaded) in Pico 100 containers and two heat-able sources in Pico 200 (TDMASn, Pegasus) and Pico 300 (InCp, Epivalence) liquid-based precursor containers. 20 nm $SnO_X$ was grown at the ALD chamber using the combination of the TDMASn and DI water precursors with a pulse time set at 1.4 and 1.6 s for TDMASn(IV) and $H_2O$, respectively, under the reactor temperature of 100 °C for 150 cycles[59].

## 7.2 Spectroscopic Ellipsometry

For the ellipsometry fitting, we combined data from Variable Angle Spectroscopic Ellipsometry (VASE) measurements with simple transmission measurements. VASE involves measuring the change in polarization of light upon reflection from the sample at multiple wavelengths and incidence angles. These polarization changes are described by the ellipsometry parameters Ψ and Δ, which represent the amplitude ratio and phase difference between the p- and s-polarized components of light, respectively. Ψ and Δ can be modelled based on the thickness and complex refractive indices of each layer in the sample stack, allowing these parameters to be determined by fitting the measured Ψ and Δ values. The addition of transmission data helps to constrain the optical model, particularly in wavelength regions with high transparency, by providing additional reference points for fitting.

Measurements were carried out using a J.A. Woollam RC2 Ellipsometer, and data analysis was performed using the CompleteEASE software provided with the instrument. For the VASE measurements, Ψ and Δ were recorded at the following angles: 55°, 60°, 65°, 70°, and 75° for the $Cs_{0.13}FA_{0.87}Pb(I_{1-x}Br_x)_3$ (x=0,0.25,0.5,0.75,1) samples; and 50°, 60°, and 70° for the $Cs_{0.1}FA_{0.61}MA_{0.3}Sn_xPb_{(1-x)}I_3$ (x=0,0.2,0.3,0.4,0.5,0.6) samples. Fewer angles were used for the mixed lead-tin samples to reduce measurement time, as these samples degrade readily in air. A stream of inert nitrogen gas was flown over the mixed lead tin samples as an additional precaution against air degradation during the measurement.

The samples consisted of multiple layers, including a glass substrate, a tin oxide ($SnO_2$) adhesion layer, and the perovskite layer. The optical constants for each layer were determined individually by sequentially measuring single-layer samples: first, the glass substrate alone to determine its optical constants; then a glass/$SnO_2$ sample to determine the $SnO_2$ values; and finally, the full perovskite samples. This stepwise approach improved the uniqueness of the final fit by ensuring that only one layer's optical constants were fitted at a time. The glass substrate was modelled using a Cauchy optical model, while the $SnO_2$ layer was modelled using a B-spline representation, where the optical constants were defined by a set of adjustable B-spline nodes. A piece of Scotch tape was applied to the back of the glass substrate before measurement to act as a scattering layer, eliminating back reflections and allowing the substrate to be modelled as infinitely thick. The fitted $SnO_2$ layer thickness was assumed to be the same across all samples and was not varied during the fitting of the perovskite layers. For the $Cs_{0.1}FA_{0.61}MA_{0.3}Sn_xPb_{(1-x)}I_3$ (x=0,0.2,0.3,0.4,0.5,0.6) samples, an additional organic monolayer was present on top of the SnO$_2$ layer. However, this layer was too thin to be detected by the measurement.

For the perovskite layers, a B-spline approach was used, allowing us to fit the dielectric constants without knowledge of the exact energies and shapes of the underlying oscillators that would make up the complex dielectric extinction. To account for vertical inhomogeneity, we applied a simple dielectric-constant grading scheme. In this scheme, the dielectric function at the top surface and the bottom interface were each allowed to vary independently. The dielectric function throughout the rest of the film was then represented as a linear interpolation between these two boundary values, providing a smooth, graded profile across the film thickness. This approach accounted for slight degradation at the film's surface due to ambient exposure. We used a simple grading method with five slices, where the optical constants at the bottom differed from the top by a fixed percentage factor[60]. After the fit, the dielectric fit values from the 'bottom' of the graded stack was chosen, as this represents the most pristine material. Surface roughness was included in the fitting, but it was not sufficient to force the extinction coefficient to zero beyond the band edge.

Initially, the transparent wavelength range (wavelengths greater than the band edge) was fitted using a Cauchy model. This served as a starting point for a B-spline expansion, where the Cauchy model was converted into a B-spline representation, and additional nodes were progressively added toward shorter wavelengths. In cases where the perovskite surface roughness made it difficult to fit the transparent region accurately, we constrained the fit to wavelengths up to just below the band edge, as data beyond the band edge would be discarded when incorporating PDS/FTPS data.

### 7.3  Fourier Transform Photocurrent Spectroscopy

Fourier-transform photocurrent spectroscopy (FTPS) was measured on a custom-built system with a Bruker Vertex 80v Fourier-transform Interferometer, a tungsten-halogen near-infrared source, a CaF$_2$ beam splitter, and a transimpedance amplifier. Perovskite solar cells with an active area of 0.25 cm$^2$ were illuminated and used as a photodetector in ambient atmosphere. External Quantum Efficiency (EQE) spectra obtained were calibrated against the spectral response of a certified Si reference solar cell from Newport. Encapsulated solar cell devices were held in open-circuit condition in air with a relative humidity of ~45%, and a minimum of four pixels were measured per perovskite composition. To increase both the sensitivity and the dynamic range (up to 6 orders of magnitude in photocurrent) of all EQE spectra below the bandgap, above-bandgap illumination were filtered out with a 780 nm long-pass colour filter (Schott RG780).

### 7.4  Photothermal Deflection Spectroscopy

The samples were placed inside a quartz cuvette filled with a thermos-optic liquid (3M Fluorinert FC-72). A halogen lamp was coupled to a 250-mm focal length grating monochromator to provide a tuneable excitation source. A mechanical chopper operating at 10 Hz modulated this beam. The PDS setup operated in a transverse configuration, where a 670 nm probe laser beam was directed parallel to the sample surface in the region illuminated by the pump beam. As the sample absorbed monochromatic light, thermal energy was released, altering the refractive index of the surrounding liquid and inducing deflection of the probe beam. This deflection was detected by a quadrant silicon photodiode and analysed synchronously using a lock-in amplifier (Stanford Research Systems SR830). By recording the laser beam deflection as a function of the excitation wavelength, the absorptance of the sample was determined, which is linearly proportional to the absorption coefficient.

## Data availability

Code and data for this paper is available in our github[55]: https://github.com/akashdasgupta/Perovskite-Dielectric-Constants-Repository